\documentclass[10pt]{article}
\usepackage{latexsym,graphicx}
\usepackage{latexsym,graphicx}
\newcommand{\be}{\begin{equation}}
\newcommand{\ee}{\end{equation}}
\def\n{\noindent}
\catcode `\@=11 \catcode `\@=12
\begin{document}
\begin{center}
\large{\bf {`Zero-spin-photon hypothesis' finds another important
application: Could possibly solve the `infinity-problem' of QED
without the need of renormalization}} \\
\vspace{10mm}
\normalsize{R. C. Gupta $^1$,  Anirudh Pradhan $^2$, V. P. Gautam $^3$,
M. S. Kalara $^4$, B. Das $^5$, Sushant Gupta $^6$} \\
\vspace{5mm}
\normalsize{$^{1}$ Institute of Technology (GLAITM), Mathura-281 406, India\\
E-mail: rcg\_iet@hotmail.com, rcgupta@glaitm.org}\\
\vspace{5mm}
\normalsize{$^{2}$ Department of Mathematics, Hindu
Post-graduate College, Zamania-232 331, Ghazipur, India \\
E-mail: acpradhan@yahoo.com, pradhan@iucaa.ernet.in}\\
\vspace{5mm}
\normalsize{$^{3}$ Theoretical Physics, IACS, Calcutta, India\\
E-mail:vpgautam@yahoo.com}\\
\vspace{5mm}
\normalsize{$^{4}$ Nuclear Engineering Dept., I.I.T. Kanpur, India\\
E-mail: msk@iitk.ac.in}\\
\vspace{5mm}
\normalsize{$^{5, 6}$ Department of Physics,Lucknow University,
Lucknow-226 007, India}\\
\normalsize{$^{5}$E-mail: bdas4569@yahoo.com}\\
\normalsize{$^{6}$ E-mail: sushant1586@gmail.com}\\
\end{center}
\vspace{10mm}
\date{July 11, 2009}
\begin{abstract}
`Zero-spin-photon hypothesis' as proposed in an earlier paper
\cite{ref1} states that: `due to inevitable consequence of the
second-law of thermodynamics and spin-conservation, the
`zero-spin-photon' is generated in pair-production process (of
elementary particles), which decays into neutrino and antineutrino'.
The zero-spin photon hypothesis explains \cite{ref1} several riddles
of physics and universe. In the present paper, it is shown that `the
zero-spin photon hypothesis' when incorporated into the higer-order
Feynman diagram (with a closed-loop) could possibly solve the
half-a-century-old and famous `infinity-problem' of QED, and thus
could avoid the need of the so called `re-normalization' procedure.
\end{abstract}
\smallskip
\n Key words: Feymman diagrams, Zero-spin-photon, Infinity problem,
Re-normalization  \\
\n PACS: 23.20.Ra, 14.60.Lm, 98.80.-k, 11.30.Er \\
\section{Introduction}
Laws of thermodynamics are universally valid. The first-law is about
the `{\it conservation} of energy' whereas the second-law is about
`{\it conversion} of energy'. Engineers give equal weightage to both
these laws, but unfortunately the second-law has largely been
ignored by physicists. The second-law of thermodynamics which
basically tells about the 'irreversibility' of the energy (heat and
work) conversion process, has far-reaching consequences [1-3]. When
examined from thermodynamics perspectives, it is found [1-3] that:
all forms of energies including mass energy (${\rm { E = mc^{2}}} $)
are `work', except the radiation energy (${\rm { E= h\nu}} $) which
is `heat'. It is also concluded therein [1-3] that in fact energy
carried by particles-with-mass are `work' whereas energy carried by
massless particle `photon' is `heat'. The important consequence is
that it necessitates generation of a residual photon in the
pair-production process (of the elementary particles); the
spin-conservation requires the residual-photon to be of spin zero;
the `zero-spin photon' being unstable further decays into a pair of
neutrino and antineutrino. It has been stressed therein [1] that any
`proposal' even as the level of hypothesis, which solves some of the
riddles, should be welcomed. The `zero-spin photon hypothesis'
\cite{ref1} seems to be corrects as it indeed explains reasonably
well several riddles of physics and universe including that of the
neutrino-handedness
and parity-violation.\\\\
Quantum Electro-dynamics (QED) has been one of the most successful
theories of twentieth century physics. However, the problem of
`infinity' had plagued the theory for couple of decades until a
stop-gap arrangement or the so called `re-normalization' procedure
was developed to get rid of the infinity-disaster. Though
successful, the `re-normalization' has been considered disturbing
since beginning (as mentioned in section-$2$). In the present paper;
an alternative way (suggested in section-$4$), to solve the
infinity-problem of QED without the need of re-normalization, is
being suggested with the help of the `zero-spin photon
hypothesis'-proposed \cite{ref1} in an earlier paper (briefly
revisited in section-$3$ of this paper).
\section{Infinity-problem, Re-normalization and its Criticism}
\subsection{Infinity-Problem}
Evaluation of the Feynman-diagrams to determine the `amplitude'
\emph{M} for the process in question is necessary for calculation of
decay rates ($ \Gamma $) and scattering cross-section ($ \sigma $),
as the case may be. When the higher-order Feynman diagrams with
closed-loops are evaluated for amplitude \emph{M}, the `integral'
diverges to infinite-value causing the infinity-problem.
Infinity-problem in Feynman-calculus seem to be the characteristic
of the closed-loops in the Feynman-diagrams. One such closed-loop
under-consideration in the present-paper is the one that arises when
a (virtual) photon emits electron and positron (pair-production)
which are soon reabsorbed (annihilation) therein.
\subsection{Re-normalization}
`Re-normalization' is a set of ad-hoc strategy and procedure to
`regularize' the integral and to absorb (cancel) the infinities with
the `re-normalized' masses and coupling-constants as effective and
running respectively. Detailed strategy, procedure and formulae can
be found in particle-physics or quantum-mechanics books
\cite{ref4,ref5}.
\subsection{Criticism of Re-normalization}
Griffith \cite{ref4} mentions about the problem of infinities in
QED/QFD while applying the Feynman-rules to the Feynman-diagrams
with closed-loops (for calculating amplitude \emph{M}) as follows:
``The `integral' is logarithmically divergent at large `q'. The
disaster in one form or the other, held up the development of
quantum electrodynamics for nearly two decades, until, through the
combined efforts of many great-physicists (critics and
supporters)--Dirac, Pauli, Kramers, Weisskopf and Bethe through
Tomonaga, Schwinger and Feynman--systematic methods were developed,
for sweeping the
infinities under the rug'', as `re-normalization'.\\\\
Dirac is reported \cite{ref4,ref6} to have critically remarked about
re-normalization as follows: ``it (re-normalization) is just a
stop-gap procedure. There must be some fundamental change in our
ideas, probably a change just as fundamental as passage from Bohr's
orbit theory to quantum-mechanics. When you get a number turning out
to be infinite which ought to be finite, you should admit that there
is something wrong with our equations, and not hope that you can get
a good theory just by doctoring up (manipulating) that number...,
with a good theory the `infinity' would never arise
in the first place''.\\\\
Kaku and Thomson \cite{ref7} reiterate as satire that ``can
`infinity minus infinity' yield a meaningful results (or in the
language of physics can  $ \infty - \infty = 0 $? Mathematically, it
is known to be indeterminate). To the critics, using one set of
infinities (arising from loops in the Feynman diagrams) to cancel
another set of infinities (arising from electric charge and mass)
looked like a parlor-trick''. They \cite{ref7} further quote Dirac
to have said on it that `` This ($ \infty - \infty = 0 $) is not a
sensible mathematics. Sensible mathematics involves neglecting a
quantity when it turns out to be small, not neglecting it (or
getting rid of it ) because you do not want it''.
\section{`Zero-spin photon hypothesis' Re-visited and using it into
single-vertex Feynman-Diagrams for (single) pair-production to yield
multi-vertices Feynman-diagram for (double) pairs-Production:}

\subsection{`Zero-spin photon hypothesis'- Revisited}
The second-law of thermodynamics, which spells of `irreversibility',
prohibits the full-conversion of heat (radiation energy ${\rm{
\gamma}} $) into work-energy (or generation of pair-particles of
energy ${\rm { 2mc^{2}}} $) and thus \cite{ref1} necessitates some
residual energy (${\rm { \gamma_{0}}} $) to come-out. As described
in the earlier- paper \cite{ref1}; the inevitable consequence of
second law of thermodynamics and spin-conservation necessitates the
generation of a zero-spin photon (${\rm { \gamma_{0}}} $) in
pair-production (of electron and positron), which (${\rm {
\gamma_{0}}} $) being unstable subsequently decays into a pair of
neutrino and antineutrino. The generation of zero-spin photon ($
{\rm {\gamma_{0}}} $) in electron-positron pair-production and its
subsequent decay into neutrino- antineutrino pair-production makes
together (combined) pairs-production process, as shown in Fig.$1$.
The strong photon ${\rm { \gamma}} $ is shown as wave with
continuous wavy-line, whereas the weak zero-spin-photon $ {\rm
{\gamma_{0}}} $, being unstable, is shown as dotted-wave. The
proposed hypothesis \cite{ref1} seems okay as it is able to explain
many riddles of physics and universe, including the famous
parity-violation \cite{ref1}.
\begin{figure}[htbp]
\centering
\includegraphics[width=13cm,height=6cm,angle=0]{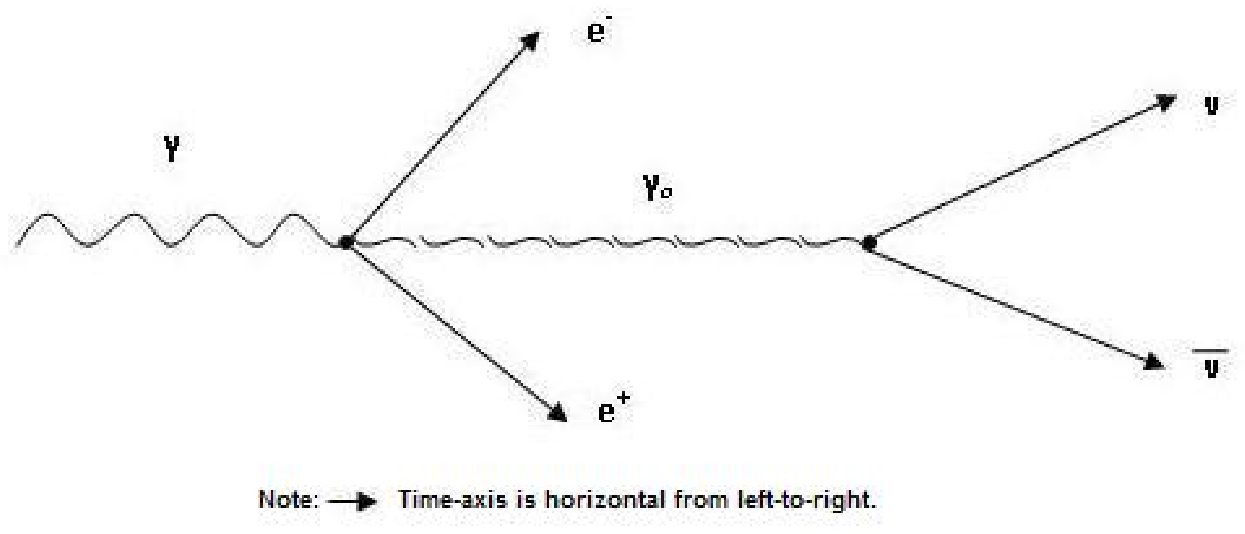}
\caption{Schematic-diagram for `zero-spin-photon hypothesis':
generation of `zero-spin-photon $\gamma_{0}$' during electron
positron pair production and its subsequent-decay into neutrino and
antineutrino, leading to pairs-production}
\end{figure}

\subsection{Introducing the zero-spin photon hypothesis concept towards
the Feynman-Diagrams}
\subsubsection{Simple Feynman-diagram(tree)of electron-positron
pair-production and that also for neutrino-antineutrino
pair-production}

 The single-vertex Feynman trees/diagrams (wherein
only 3 lines are permissible) for electron-positron pair-production
(from the energetic gamma-ray photon $ {\rm {\gamma}} $) and that
also for neutrino-antineutrino pair production (from the residual
zero-spin photon ${\rm { \gamma_{0}}} $) are `separately' shown in
Figs.${\rm {2.a}}$ and ${\rm {2.b}}$, note that the time-axis is
horizontal from left-to-right; thus the particles traveling from
right to left towards vertex are the corresponding anti-particles.
\begin{figure}[htbp]
\centering
\includegraphics[width=13cm,height=6cm,angle=0]{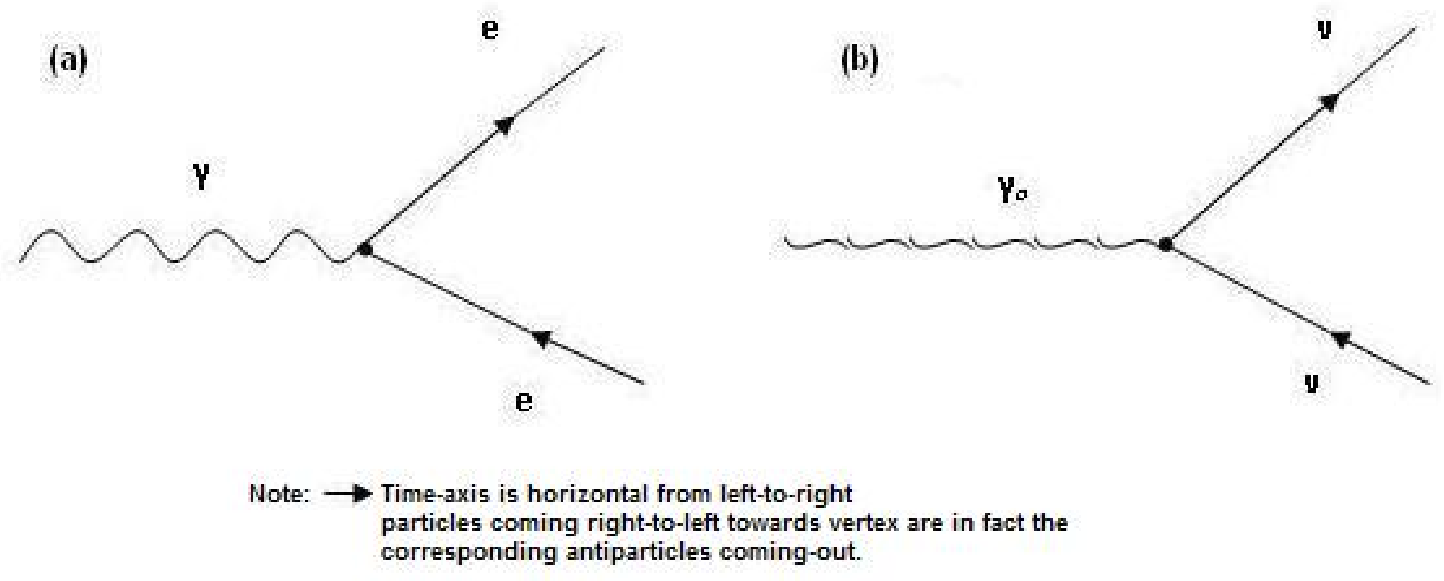}
\caption{(a) Feynman diagram for electron-positron pair-production
from $\gamma$ (b) Feynman diagram  for neutrino-antineutrino
pair-production from $\gamma_{0}$}
\end{figure}
\subsubsection{`Single-vertex Feynman diagrams (trees) for the two
pair-production process' together to yield Feynman-diagrams for
pairs-production with three or five vertices}
\begin{figure}[htbp]
\centering
\includegraphics[width=12cm,height=10cm,angle=0]{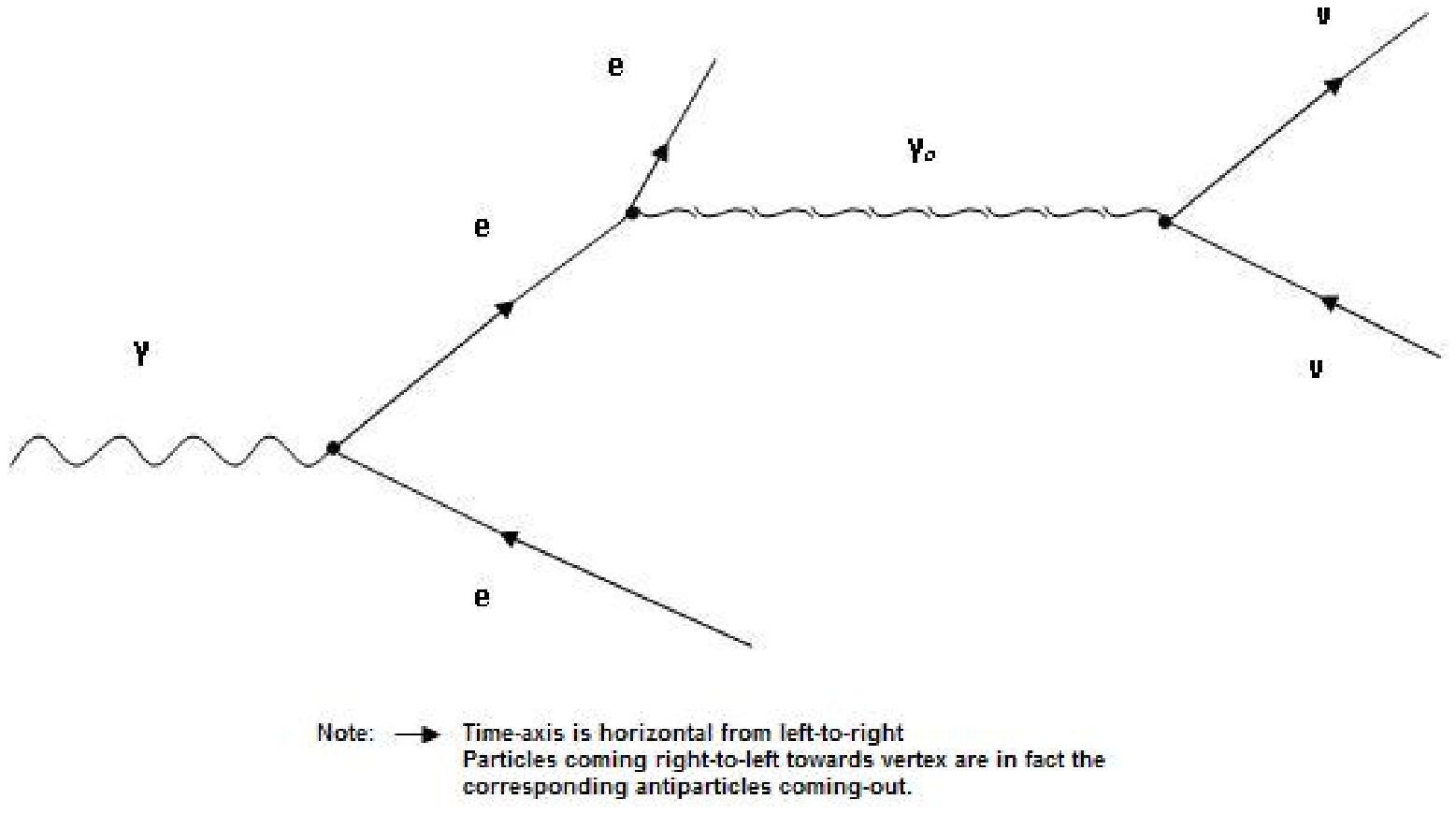}
\caption{Suggested Feynman-diagram for pairs-production with one
$\gamma_{0}$ and three-vertices}
\end{figure}
\begin{figure}[htbp]
\centering
\includegraphics[width=13cm,height=10cm,angle=0]{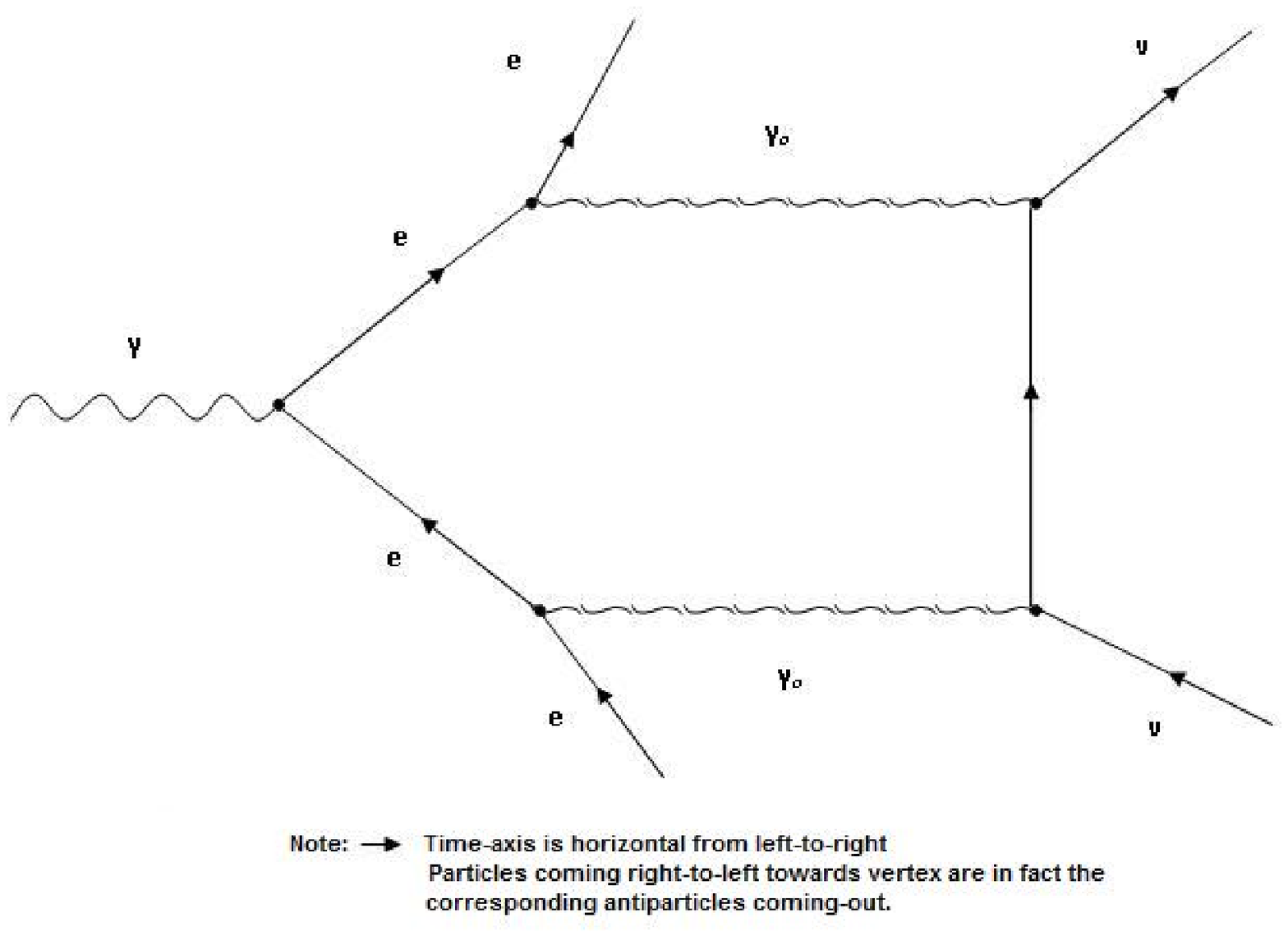}
\caption{Suggested Feynman-diagram for pairs-production with two
$\gamma_{0}$ and five-vertices}
\end{figure}
The two pair-production process (as mentioned in Section $3.2.1$ and
shown in Figs.${\rm {2.a}}$ and ${\rm {2.b}}$) when
combined-together is named (plural) as the pairs-production, and is
mentioned earlier (in section $3.1$) and is shown schematically in
Fig.$1$ ( The Fig.$1$, at the first vertex, contains `4' line(s),
which is not permissible in such Feynman-diagram). However, it can
also be shown compatible (permissible) as Feynman-diagrams as in
Figs.$3$ (with three vertices) and in Fig.$4$ (with five vertices).
Pairs-production can possibly occur in either way; Fig.$4$ is more
likely, being symmetrical. Herein it is assumed that the
uncertainty-principle ${\rm { \Delta {E} \Delta {t} =
\frac{\hbar}{2}}}$ (which permits violation of `energy-conservation
or the first law of thermodynamics' for a moment) also permits
violation of second-law of thermodynamics as well, thus the $ {\rm
{\gamma_{0}}} $ can come-out after a moment from next-vertex to be
in accordance with Feynman-diagram, with only 3-lines on each
vertex.

\section{Introducing the `zero-spin photon hypothesis' onto the closed-loop of
Feynman-diagram (for Moller-scattering) as possible solution to the
infinity-problem of QED}
\subsection{Closed-loop in Feynman-diagrams as source of the infinity-problem}
Infinity-problem (divergence of the `integral' for amplitude
\emph{M}) normally arises due the closed-loops in the higher-order
Feynman-diagrams, such a closed-loop situation for the
electron-electron Moller-scattering is shown in a conventional-way
in Fig.$5$. The closed-loop comes due to `pair-production' and its
subsequent `annihilation' therein.\\\\
To get rid of this infinity-problem, a set of procedure called
`re-normalization' is adopted. The `re-normalization' though
successful, seems dubious and has been criticized enough. The
half-a-century-old problem of infinity has not been solved as yet,
but rather it has only been `managed' through `re-normalization'.
The present paper is an honest attempt which could possibly solve
the infinity-problem without re-normalization, if the `zero-spin
photon hypothesis' is incorporated into the closed loop of
higher-order Feynman-diagram, described as follows.
\begin{figure}[htbp]
\centering
\includegraphics[width=8cm,height=10cm,angle=0]{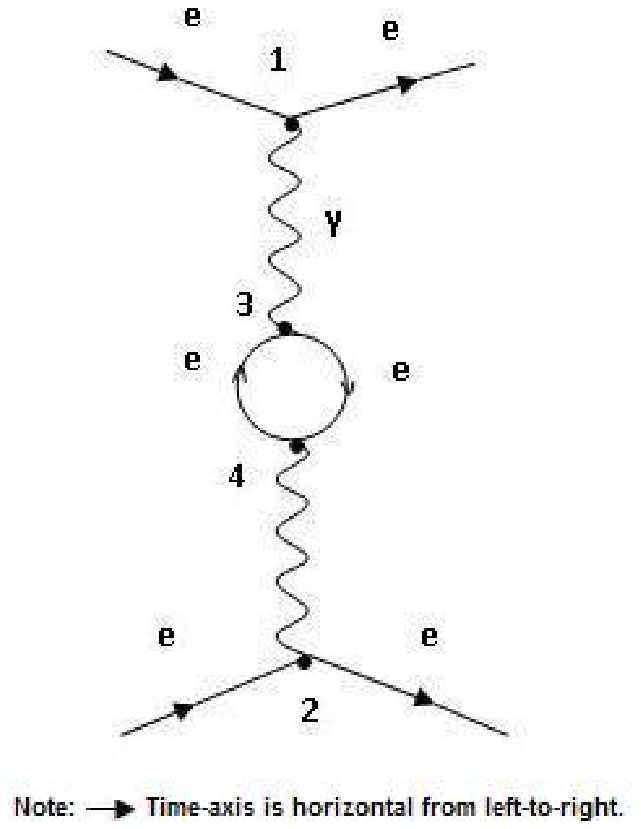}
\caption{Higher-order Feynman-diagram for electron-electron Moller
scattering (with four-vertices) with the smooth closed-loop in it,
which leads to the `infinity-problem"}
\end{figure}

\subsection{Introducing the `zero-spin photon and its subsequent decay'
onto the closed-loop of the Feynman-diagram -- modifications in the
Feynman-diagram--could possibly solve the infinity-problem}

As discussed in the `zero-spin photon hypothesis' \cite{ref1},
revisited here in section-$3.1$; electron-positron pair-production
from a photon (${\rm { \gamma}} $) also necessitates generation of
zero-spin photon ($ {\rm {\gamma_{0}}} $) which subsequently decays
as pair-production of neutrino and antineutrino, making the whole
process as `two' pair production(s) referred simply as
`pairs-production'. The Feynman diagrams(Figs.$3$ and $4$) of
pairs-production with additional vertices could now be
`superimposed' onto the closed-loop of the
Moller-scattering (Fig.$5$).\\\\
Thus two possible (modified) Feynman-diagrams are produced with six
or eight vertices as shown in Figs.$6$ and $7$ respectively.It may
be seen and noted (and idiomatically said) that the closed-loop (the
problematic viscous-circle) is no longer smooth-curve but
broken-down(in pieces) and therefore the root-cause of the
infinity-problem is broken (the possible `technical' explanation for
why the infinity-problem goes off  is briefly mentioned
in the next paragraph).\\\\
\begin{figure}[htbp]
\centering
\includegraphics[width=13cm,height=13cm,angle=0]{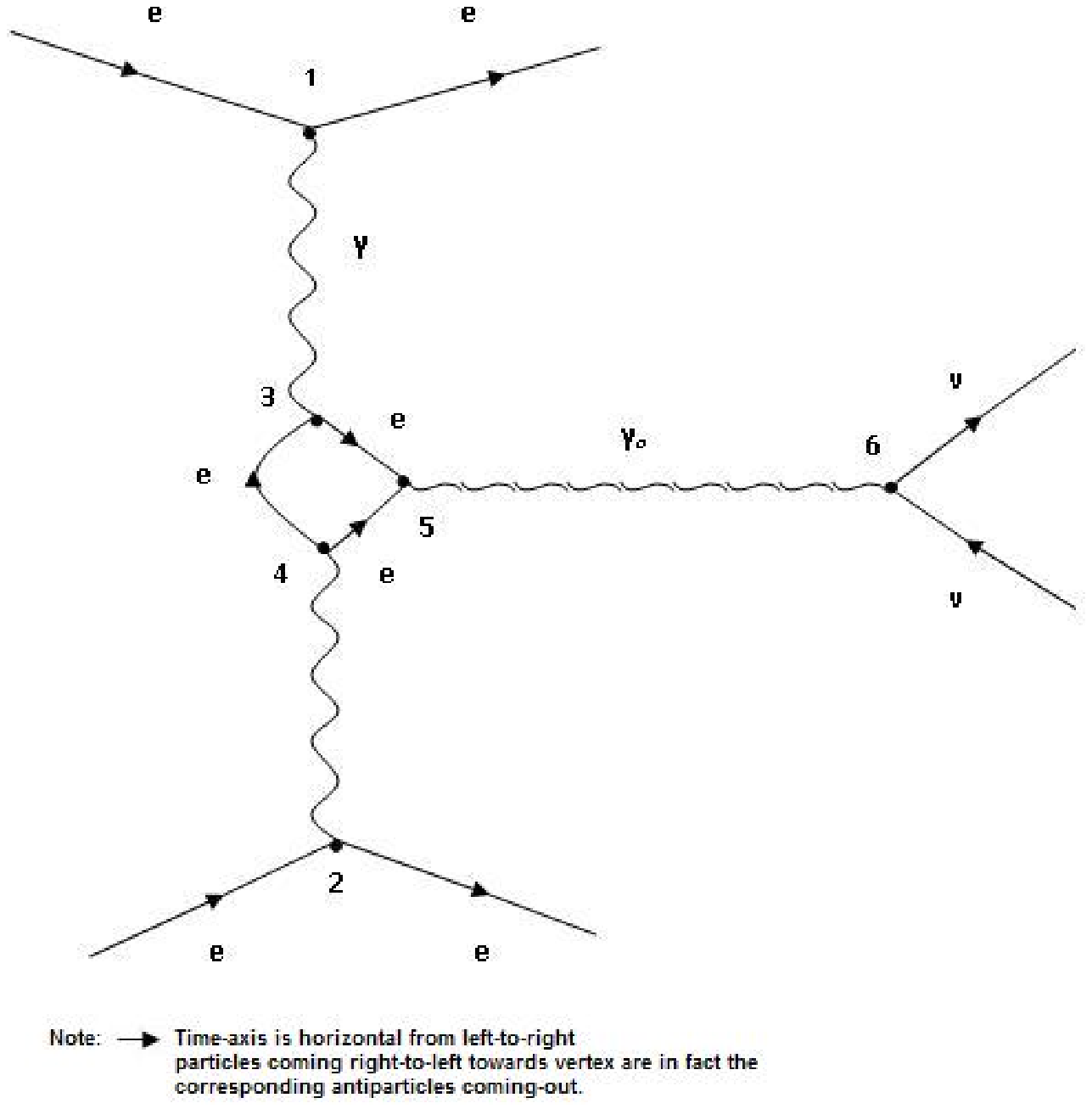}
\caption{Modified higher-order Feynman-diagram for the
Moller-scattering (with six vertices) using one `zero-spin-photon',
super imposing figure $3$ on $5$}
\end{figure}
\begin{figure}[htbp]
\centering
\includegraphics[width=10cm,height=13cm,angle=0]{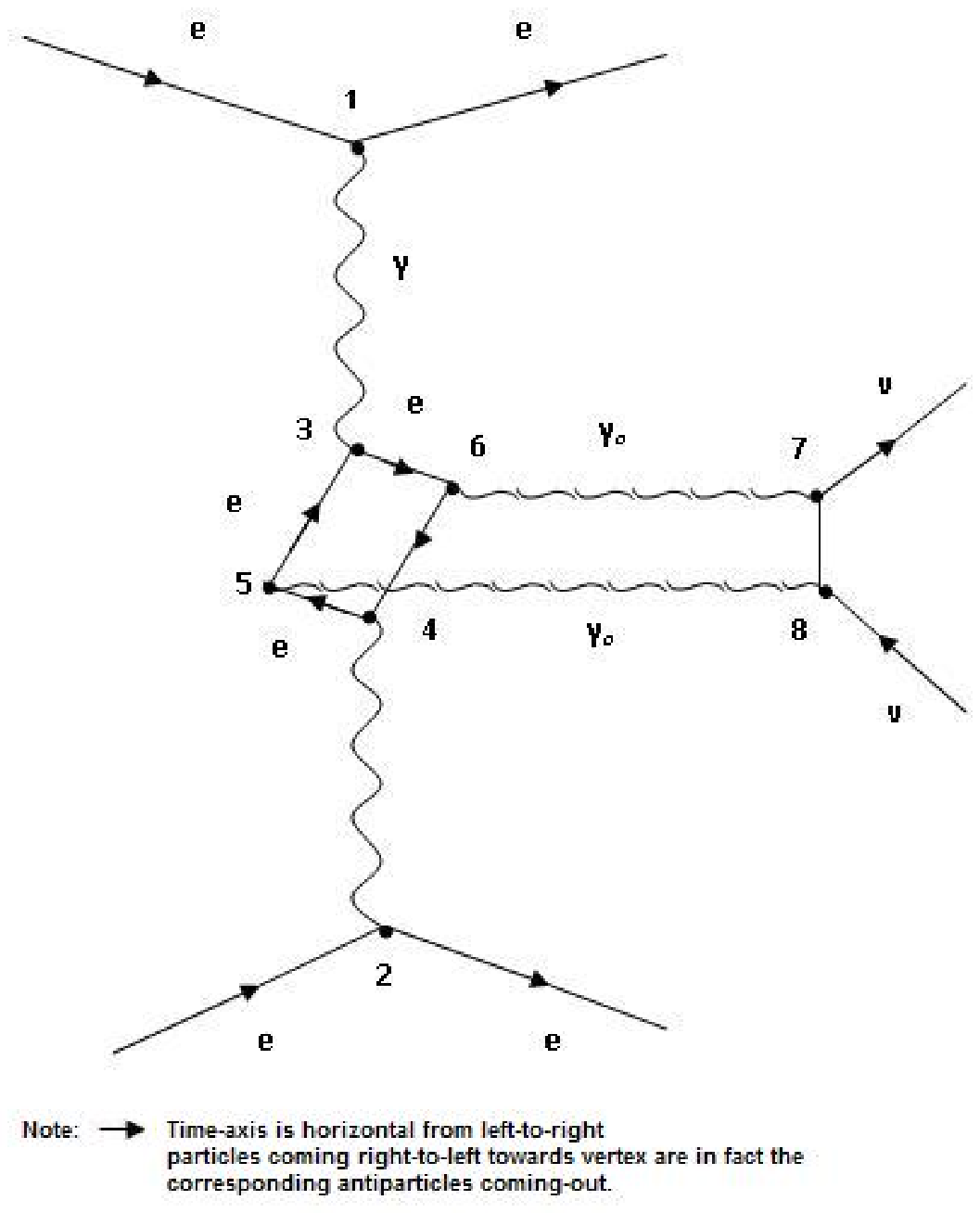}
\caption{Modified higher-order Feynman-diagram for the
Moller-scattering (with eight vertices) using two
`zero-spin-photon', super imposing figure $4$ on $5$ }
\end{figure}
Due to more internal-line(s) (as in the Figs.$6$ and $7$) more
contributions will be there in the `integrand' through
`propagators', thus having more terms (of internal four-momentum
`q') in the denominator, it is therefore expected that the
`integral' instead of being divergent would rather now be convergent
(possibly inverse dependence). This could thus solve the infamous
infinity-problem altogether, making the need of `re-normalization'
unnecessary or redundant. Incorporating `zero-spin photon
hypothesis' onto the conventional Feynman diagram seems to make the
diagram complex but solution cleaner. The zero-spin-photon, being
unstable (thus shown as dotted-wavy-line) comes only for a short
while to decay into the realistic neutrino and antineutrino; thus
from the internal-line-propagator point-of-view the wavy-lines may
possibly
be considered absent/omitted.\\\\
The authors invite the physicists of the world to tackle the
modified Feynman-diagrams (Fig.$6$ and $7$) rigorously. The
`zero-spin-photon'-path is laid down, directions for solution are
indicated. It is to be seen who wins the race to be the first to
`actually' solve the `infinity-problem', of say, Moller-scattering,
without using the crutches of `re-normalization'? Also to be seen
that how the two other types of loops in the Feynman-diagrams (those
not discussed at all in the present-paper) are to be tackled and who
does that? In addition; the `zero-spin photon' - approach, for
avoidance of infinity, may also be applied to other
scattering-process and decay-process.
\section{Discussions}
`Law(s) of thermodynamics' are not simply meant for engineers only.
Its greatness goes much beyond \cite{ref8,ref9}. It is said
\cite{ref9} `that the four laws of thermodynamics (Zeroth, First,
Second and Third) drives the universe, and that not knowing
(appreciating) the second-law of thermodynamics is like never having
read a work of Shakespeare'! Interestingly, it has been shown [1-3]
that though remotely distinct, `second-law of thermodynamics' and
`special-relativity' are linked to each other. Physicists have great
respect for special-relativity and other physical-laws, but
in-general ignore the importance of the second-law of
thermodynamics. Anyway, the `zero-spin photon ${\rm{\gamma_{0}}} $'
is much smaller than $ {\rm {\gamma}} $; so should we bother about
including ${\rm {\gamma_{0}}} $ ? It is true that, `though most of
small differences don't make much of a difference, but even if the
small difference pops up at crucial point then it can make all the
difference' \cite{ref10}. Ignoring the second-law of thermodynamics
is the basic etiology of the infinity-problem, which can be overcome
if the zero-spin-photon (which is the outcome of the second-law of
thermodynamics) is incorporated in the Feynman-diagram.\\\\
The root cause of the infinity-problem is known to be the
closed-loops in the higher-order Feynman-diagrams. One such case
considered in this paper is the Moller-scattering (Fig.$5$) wherein
the exchange-photon (${\rm { \gamma }}$) creates the
electron-positron (pair-production) and its subsequent recombination
(annihilation) results in a closed-loop in the diagram.
`Annihilation' is thermodynamically okay; but the simple (single)
pair-production is, in a way, thermodynamically wrong or incomplete,
it should be rather be pairs-production as mentioned in section
$3.1$. Once the inclusion of `zero-spin photon' is made (as in
modified Feynman-diagrams, Figs.$6$ and $7$) to incorporate the
correct pair(s) - production, the problem of infinity should go away
because the troublesome vicious-circle (the smooth continuous
closed-loop of the original Feynman-diagram Fig.$5$) is broken-down
(in pieces, as shown in the modified Feynman-diagrams Figs.$6$ and
$7$) to first-order (slope) discontinuity. A few additional-vertices
and internal-lines are thus introduced in the modified
Feynman-diagram(s), which would contribute via `propagators'
possibly in such a way to have more powers of `q' in the denominator
of the amplitude's `integrand' ultimately making the `integral'
inversely convergent. The ill-behaved original integral
should now become well behaved!\\\\
It is anticipated in the present paper, that the problem of infinity
originating out of the smooth closed-loop (as in Fig.$5$) can most
likely be solved with the incorporation of `zero-spin photon'
hypotheses. The authors ask the physicists and graduated-students to
analyze the new suggested (modified) Feynman-diagrams with the
broken-down loop (Figs.$6$ and $7$); the authors have indicated in
the previous paragraph(s)-- how to possibly get the `integral'
convergent thus avoid the infinity-problem altogether. But this is
not the end of story. There are other two more types of loops which
too cause divergence problem but these have neither been tackled nor
answered in the present-paper. But at least, a new light of
`zero-spin-photon' has been thrown-upon possibly in the
right-direction which can illuminate the path.

\section{Conclusions}
Normally physicists bother much about momentum-conservation and
energy-conservation (`first-law' of thermodynamics), and that is
what the Feynman-calculus insures; but the `second-law' of
thermodynamics has completely been ignored. This non-consideration
of second-law of thermodynamics is basically the origin and
root-cause of the infinity-problem in QED; therefore to avoid this
disastrous problem, the right perspective of second-law of
thermodynamics should be incorporated via `zero-spin photon
hypothesis', and that is what is done in the present paper by
modifying the Feynman-diagram accordingly. The herein proposed
theory/approach seems to be in the right direction because it is in
accordance with what Dirac has said \cite{ref6} that `with a good
theory the infinity would never arise in the first place'. It
appears that the half-a-century-old infinity problem, which so far
has only deem managed with re-normalization, can possibly be solved
without re-normalization. The authors anticipate a wide-scale
repercussions of this paper, as the suggested approach is expected
to make changes in the shape and fate of the Feynman-diagrams and
the particle-physics.

\section*{Acknowledgements}
Two of the authors (A. Pradhan and Sushant Gupta) thank the
Inter-University Centre for Astronomy and Astrophysics, Pune, India
for providing facility where part of this work was carried out. The
author (R. C. Gupta) thanks IET/UPTU, Lucknow and GLAITM, Mathura
for providing facility for the research.

\end{document}